# FRACTIONAL ORDER DERIVATIVE APPROACH OF VISCOELASTIC BEHAVIOR OF TROPICAL WOOD

**Loic Chrislin NGUEDJIO[1,2], Rostand MOUTOU PITTI[2,3], Benoit BLAYSAT[2], Pierre Kisito TALLA[1] , Nicolas SAUVAT[2], Joseph GRIL[2]**

**ABSTRACT:** For some time now, wood has offered itself as an alternative to other modern construction materials, and has become the material of choice for structures, mainly because of its renewable nature, durability and ease of shaping. However, once in service, even at room temperature and under low stresses, it deforms and faces the problems of creep and recovery. The aims of this work is to model and predict the viscoelastic deformations of tropical wood by a rheological approach based on fractional calculus theory. Frist, Zener fractional model was used to elucidate these phenomena. The simulations show that the proposed model fit the creep experimental data with an average reliability of 96% and the recovery process with a reliability of 99%. The optimal parameters of this model, determined through an optimization algorithm, exhibit sensitivity to the stress level. To address this issue and enhance the predictive capability of the model, nonlinearities are incorporated into the fractional model, resulting in modified versions that remain applicable across various stress levels.

## 1 – INTRODUCTION

In recent years, the demand for wood materials in construction industry has grown considerably. Thanks to its low density, the wood material offers a better strength-weight ratio and its cellular structure makes it light, thus reducing the overall mass of the structure while keeping it very solid. According to a study carried out by the Passive House institute in 2015, 55% of buildings are certified to be built in wood material [1], which demonstrates the great potential of this material by giving it the place of best candidate against its competitors due to its mechanical, thermal and ecological qualities. Once in service in a structure, wood material is subject to a set of constraints (environmental constraints, external and internal loads, etc.) which push it to exhibit faulty behaviors such as creep and recovery [2, 3], that cause damage , progressive ruin and breakage of the structural element. It is becoming urgent to find ways and means to control these harmful behaviors in order to guarantee the safety of wooden structures. Tropical woods are also affected by the issue of long-term deformation. Although numerous studies have already characterized their physical and mechanical properties [4-6], the need to model their viscoelastic behavior remains essential [7].

Rheological modeling is considered today as a solution that provides satisfactory results [8]. Fractional rheological models offer the advantage of preserving the material's memory effect by capturing its behavioral history while significantly reducing the number of model parameters, thereby lowering simulation and design costs [9]. Recently, Nguedjio et al. [10] and Atchounga et al. [11] have employed this category of models to describe the creep and recovery behavior of two tropical wood species. A comparison with classical rheological models reveals the superior performance of fractional rheological models. Furthermore, additional studies using these models corroborate these findings [12-14].

A key observation across these studies is that the model parameters are influenced by the stress level, which compromises their predictive capability. Ideally, a truly predictive model should maintain consistent parameters regardless of the applied stress. To address this sensitivity, nonlinearities whether exponential, logarithmic, or linear are often introduced into the model [15-17].

This paper aims first to determine the optimal parameters of the constant-order fractional Zener model by comparing it with experimental creep-recovery data obtained through four-point bending, using the Levenberg-Marquardt algorithm. Next, it seeks to introduce an appropriate nonlinear formulation to develop a truly predictive model.

---

[1] Loic Chrislin NGUEDJIO, Pierre Kisito TALLA, Mechanics and Modeling of Physical Systems Research Unit (UR-2MSP), Department of Physics, University of Dschang, BP 067, Dschang, Cameroon, lchristlin2000@gmail.com, tpierrekisito@yahoo.com

[2] Loic Chrislin NGUEDJIO, Benoit BLAYSAT, Joseph GRIL, Nicolas SAUVAT, Rostand MOUTOU PITTI, Clermont Auvergne University, INP, CNRS, Clermont-Ferrand, F-63000, France, lchristlin2000@gmail.com, rostand.moutou_pitti@uca.fr, joseph.gril@cnrs.fr, benoit.blaysat@uca.fr, nicolas.sauvat@uca.fr,

[3]Rostand Moutou Pitti, CENAREST, IRT, BP 14070, Libreville, Gabon, rostand.moutou_pitti@uca.fr

Following this introduction, the paper proceeds with a presentation of the mathematical model, a description of the material and experimental protocol, and an analysis and discussion of the results. The paper concludes with final remarks and perspectives for future research.

## 2 – BACKGROUND

This section presents the fractional rheological model and numerical method used in this work to calculate the relevant parameters of the model.

### 2.1 FRACTIONAL ZENER MODEL

To form the Zener fractional model, a spring is connected in serie with a basic fractional Kelvin model (Fig.1) [7]. The fractional Kelvin-Voigt model is derived by replacing the dashpot in its classical counterpart with an element called spring-pot, which imparts the fractional nature to the model. The behavior of this element is governed by a law expressed in terms of the fractional derivative, as shown in Eq. 1:

$$\sigma(t) = v.D^\beta\big(\varepsilon(t)\big), \qquad 0 < \beta < 1 \qquad (1)$$

where $\sigma(t)$ represents the stress history, $\varepsilon(t)$ the strain history, $v$ the fractional viscosity, $D^\beta$ the fractional derivative and $\beta$ the fractional order.

An electromechanical analogy analysis of the model in Fig. 1 yields the following differential equation describing its temporal behavior:

$$E_0.D^\beta\big(\varepsilon(t)\big) + b.\varepsilon(t) = \qquad (2)$$
$$D^\beta\big(\sigma(t)\big) + a.\sigma(t),$$

where $a = \frac{E_0 + E_1}{v}$ and $b = \frac{E_0 E_1}{v}$. $E_0$ and $E_1$ are the elastic moduli of the model.

Under constant stress history, the solution of the fractional differential equation (Eq. 2), expressed in terms of compliance and derived using the Laplace transform method, is given by the following equation :

$$J(t) = J_0 + J_1\left(1 - E_{\alpha,1}\left(-\frac{E_1}{v}t^\beta\right)\right) \qquad (3)$$

where $E_{\alpha,\vartheta}(.)$ represents de Mittag-Leffler function [18] and $J_0 = \frac{1}{E_0}$, $J_1 = \frac{1}{E_1}$.

### 2.2 VISCOELASTIC BEHAVIOR

This paper investigates the viscoelastic behavior of creep-recovery. Initially, the material is subjected to a constant stress $\sigma_0$, maintained over time for a duration $t_c$, representing the creep phase. Upon reaching $t_c$, the stress is spontaneously removed, and the material transitions into the recovery phase. The stress state at each moment during this viscoelastic process is expressed as follows:

$$\sigma(t) = \begin{cases} \sigma_0 H(t) & \text{if } t < t_c \\ \sigma_0\big(H(t) - H(t - t_c)\big) & \text{if } t > t_c \end{cases} \qquad (4)$$

H(t) is the Headviside function.

The viscoelastic response to the applied stress is described by the following system of equations:

$$\varepsilon(t) = \begin{cases} \sigma_0 J(t) & \text{if } t < t_c \\ \sigma_0\big(J(t) - J(t - t_c)\big) & \text{if } t > t_c \end{cases} \qquad (5)$$

with J(t) is the compliance function of the model defined by Eq. 3.

### 2.3 PARAMETERS DETERMINATION

The optimal model parameters are identified by fitting the experimental data (detailed in Section 4) to the model using the Levenberg-Marquardt optimization algorithm. The numerical scheme for computing these parameters in the MATLAB environment is illustrated in Fig. 2.

## 3 – PROJECT DESCRIPTION

The plant material used in this study is the tropical wood species Sapelli (*Entandrophragma cylindricum*). Characterized by a straight, cylindrical trunk (see Fig. 3), the Sapelli tree is commonly employed in structural and shipbuilding applications. The samples analyzed were sourced from the tropical forests of southern Cameroon. The samples were cut to dimensions of 360 mm in length and 20 mm × 20 mm in cross-section, then conditioned in an environment with a relative humidity of 65% ± 5% and an ambient temperature of 20°C ± 2°.

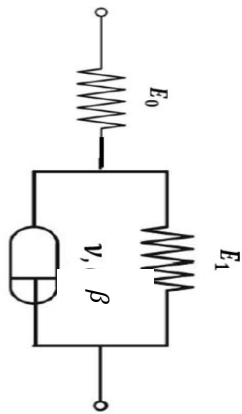

*Figure 1. Fractional Zener model.*

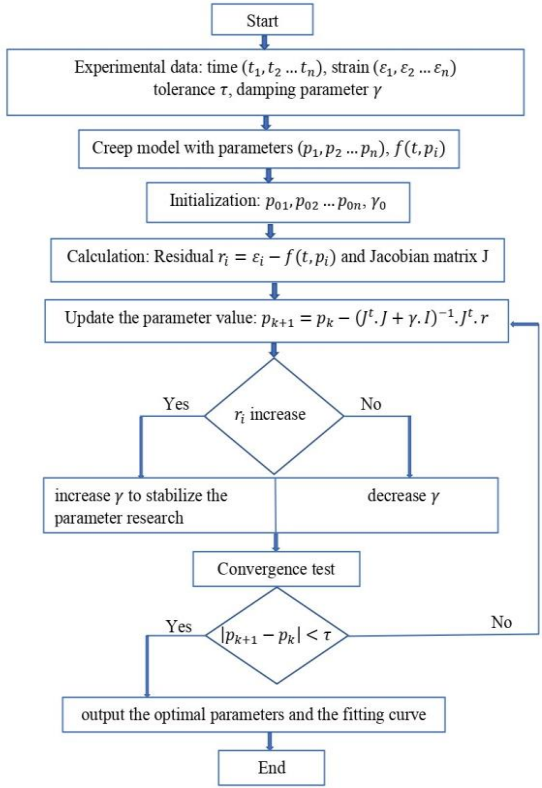

*Figure 2. Levenberg-Marquardt algorithm.*

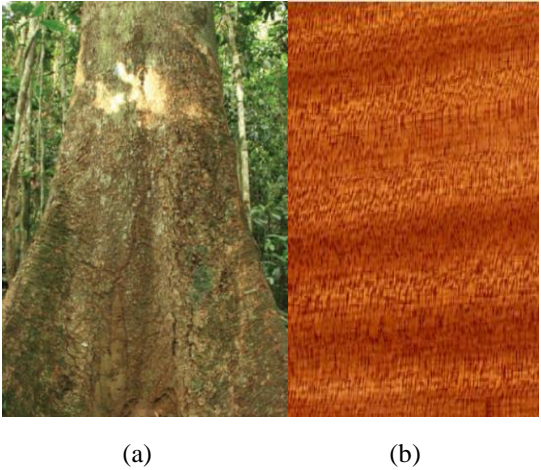

(a)                  (b)

*Figure 3. Tropical Sapelli wood : (a) Trunk, (b) Plane section.*

## 4 – EXPERIMENTAL SETUP

In this work, experimental wood deformation data are obtained through a four-point bending test (Fig. 4). Following the French Norm NF B 51-016, the tests involve placing a specimen on two simple supports and subjecting it to two loads concentrated at these supports, each applied at an equal distance from them. Initially, a constant load is maintained, and data regarding creep are gathered. Subsequently, once the creep duration has passed, the load is abruptly withdrawn, and the extensometer records deformations, reflecting the material's recovery process.

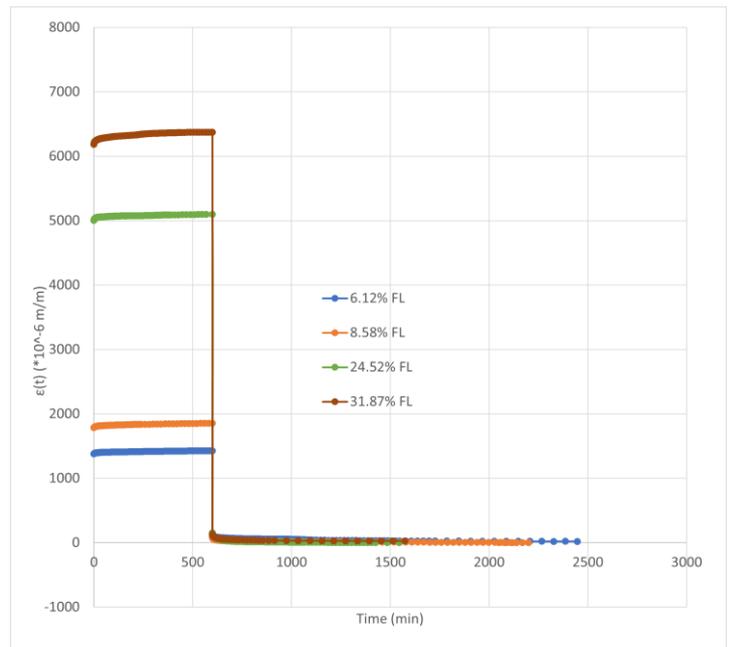

*Figure 4. Experimental setup.*

The maximum stress σ applied in the central part of the specimen is calculated from the pressure value indicated by the pressure gauge, following the relationship given in Eq. 6 :

$$\sigma(t) = \frac{3\,d\,P(t)}{bh^2} \qquad (6)$$

d is the distance between outer and inner loading points, P the applied load, b and h the width and height of the specimen, respectively.

## 5 – RESULTS

### 5.1 EXPERIMENTAL VISCOELASTIC CURVES

At the end of the experimental campaign, the collected data enabled the representation of the complete viscoelastic creep recovery behavior over time through a deformation curve (see Fig. 5).

*Figure 5. Experimental viscoelastic behavior of Sapelli wood.*

This analysis was conducted for four specimen groups, each subjected to four-point static bending under loads below the failure load (FL), specifically at 6.12%, 8.58%, 24.52%, and 31.87% of FL. The creep phase under constant loading lasts approximately tc = 600 minutes, while the recovery phase extends up to t = 2400 minutes for the specimen subjected to a load of 6.12% FL. Fig. 5 also reveals that the strain rate increases with the loading level, reflecting the material's flexibility. This suggests that higher loading levels could lead to the onset of plastic deformations.

## 5.2 MODELING VISCOELASTIC BEHAVIOR THROUGH FRACTIONAL ZENER MODEL

Once the experimental data are obtained, it is necessary to determine the optimal model parameters that best fit these data. This step enables us to assess the reliability of the proposed fractional model in simulating the viscoelastic behavior of tropical Sapelli wood. The optimization process follows the algorithm outlined in Fig. 2, using the objective function defined by Eq. 5 to capture the complete creep-recovery behavior. After optimization, the experimental curves align with the model predictions, as illustrated in Figs. 6 and 7 for creep and recovery, respectively, for the specimen subjected to 24.52% FL.

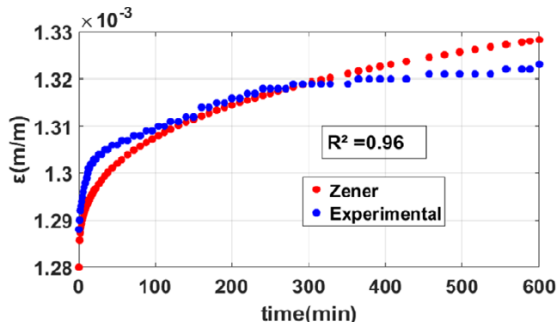

*Figure 6. Modeling creep by Zener fractional model.*

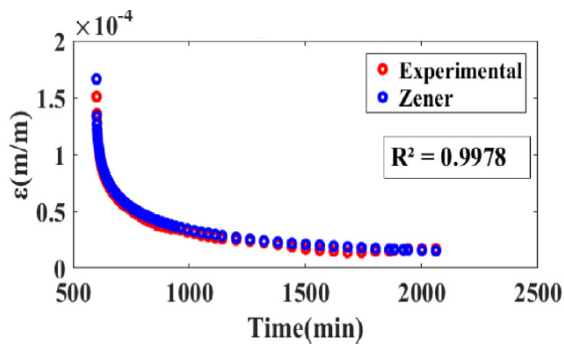

*Figure 7. Modeling recovery by Zener fractional model.*

Initial observations indicate that the Zener fractional model provides a good fit to the experimental data, particularly during the recovery phase. However, capturing the creep behavior remains challenging, especially in the initial phase within the first 100 minutes after load application. These discrepancies may stem from numerical modeling limitations or experimental uncertainties related to handling and testing conditions.

In addition to the memory effect inherent in fractional models, which enhances their ability to represent the viscoelastic behavior of materials like wood, the $R^2$ coefficient serves as a key performance indicator. The model demonstrates better accuracy in the recovery phase, achieving an $R^2$ value of 99%, while creep behavior is represented with an accuracy of 96%.

Another objective of the optimization is to determine the optimal model parameters that can be used to predict behavior under different loading histories. To this end, Tables 1 and 2 present the parameters for the creep and recovery models, respectively. The stress values σ (MPa) are computed based on the loading level using the formula given in Eq. 6.

*Table 1: Creep parameter of Zener fractional model*

| σ (MPa) | $E_0$ ($10^3$ Mpa) | $E_1$ ($10^5$ Mpa) | $\upsilon$ ($10^6$ Mpa.$\min^\beta$) | $\beta$ |
|---|---|---|---|---|
| 7.4 | 5.73 | 2.79 | 1.05 | 0.45 |
| 10.31 | 5.25 | 1.47 | 0.56 | 0.44 |
| 25.0 | 5.21 | 1.80 | 0.58 | 0.37 |
| 29.4 | 5.31 | 1.40 | 1.28 | 0.25 |

*Table 2: Recovery parameter of Zener fractional model*

| σ (MPa) | $E_1$ ($10^5$ Mpa) | $\upsilon$ ($10^6$ Mpa.$\min^\beta$) | $\beta$ |
|---|---|---|---|
| 7.4 | 6.18 | 1.23 | 0.39 |
| 10.31 | 3.33 | 0.66 | 0.29 |
| 25.0 | 3.23 | 0.72 | 0.25 |
| 29.4 | 3.80 | 0.54 | 0.18 |

The analysis of Table 1 reveals that the modulus $E_0$, which represents the material's instantaneous response to loading, remains nearly unchanged despite variations in stress. This suggests that $E_0$ can be associated with the material's intrinsic stiffness. In contrast, the other parameters exhibit significant variations with stress. The fractional order decreases as stress increases, indicating greater energy dissipation under higher loads, given the relationship between the fractional order and material stiffness [10]. Moreover, the modulus $E_1$ and the fractional viscosity $v$ show irregular variations with stress, potentially indicating nonlinearities in the material's behavior. Indeed, similar disparities have been reported by several authors in their modeling studies of the viscoelastic behavior of wood and its derivatives [11, 19, 20].

The observed disparities indicate that the fractional Zener model is sensitive to the stress level, which impacts its predictive capability. To ensure reliable simulations under different loading conditions, the model parameters should be minimally or not at all affected by stress variations. The next subsection aims to address this issue by modifying the initial model.

### 5.3 MODIFIED FRACTIONAL ZENER MODEL

To propose an adjustment to the previous model, it is essential to understand how its parameters are influenced by the stress level. In this regard, Figs. 8 and 9 illustrate the sensitivities of the parameters $E_1$ and $v$ to the stress level for creep and recovery, respectively.

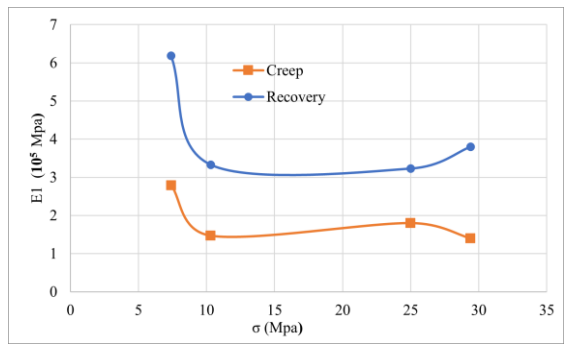

*Figure 8. Evolution of $E_1$ as a function of stress.*

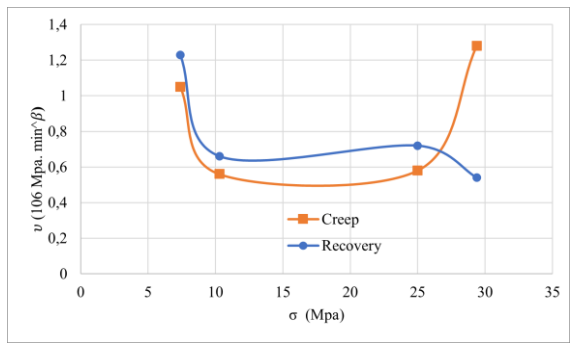

*Figure 9. Evolution of $v$ as a function of stress.*

The evolution of $E_1$ and $v$ as a function of the stress reveals the presence of two local maxima, indicating that these variations can be effectively modeled using third-degree polynomials of the form:

$$E_1(\sigma) = v(\sigma) = a\sigma^3 + b\sigma^2 + c\sigma + d \qquad (7)$$

The drawback of this approach lies in the significant increase in the number of model parameters when adopting such an evolution law. To align with the ideal of maintaining good model performance with fewer parameters, the literature generally favors limiting this type of evolution to a maximum of two parameters, ensuring that only one additional parameter is introduced compared to the initial model [17, 21].

The objective of this subsection is to identify an evolution function that closely approximates the previous polynomial function. A power function, as described in Eq. 8, appears to be more suitable, as suggested by the findings of Shimazaki et al. [17].

$$E_1(\sigma) = v(\sigma) = (\sigma + a)^b \qquad (8)$$

It is important to note that only $E_1$ or $v$ will be selected to prevent a substantial increase in the number of parameters. Consequently, the newly modified fractional Zener model, expressed in terms of the compliance function, adopts either the form of Eq. 9 or Eq. 10. These formulations are derived from the initial expression presented in Eq. 3.

$$J(t) = J_0 + (\sigma + a)^{-b}\left(1 - E_{\alpha,1}\left(-\frac{(\sigma+a)^b}{v}t^\beta\right)\right) \qquad (9)$$

$$J(t) = J_0 + J_1\left(1 - E_{\alpha,1}\left(-\frac{E_1}{(\sigma+a)^b}t^\beta\right)\right) \qquad (10)$$

In either case, we will have moved from a four-parameter fractional model to a five-parameter fractional model, with the advantage in the latter model of taking into account the non-linearity due to the variability of the wood and thus increasing the predictive capacity of the model.

## 6 – CONCLUSION

This paper introduces a modeling approach for the viscoelastic behavior of wood based on the mathematical framework of fractional calculus. The results demonstrate that this model effectively captures the creep and recovery of tropical Sapelli wood, benefiting from its incorporation of memory effects and fewer parameters compared to traditional rheological models with integer derivatives. Specifically, the creep behavior of Sapelli was simulated using the four-parameter Zener fractional model, achieving an accuracy of 96%, while the recovery was modeled with 99% accuracy under laboratory conditions. However, the parameter determination for this model revealed a certain dependence on the applied stress, complicating predictions under different loading conditions. To address this, the paper proposes a modified analytical form of the Zener fractional model that accounts for the sensitivity of the original model's parameters to stress levels. The resulting model, while introducing an additional parameter (making it five in total), promises improved predictive accuracy. Future research will focus on refining this model to identify its new parameters, ultimately creating a truly predictive model.

# 7 – ACKNOWLEDGMENTS

The authors wish to express their gratitude to the French government for its support of this work through the Eiffel Excellence program and the DIAMWOOD ANR PRCE N° 23-CE22-0006-03 project.

# 8 – REFERENCES